\documentclass{PoS}

\usepackage{graphicx}           
\input{epsf.sty}                        
\input{psfig.sty}                       

\def \apj {ApJ}

\def \aap {A\&A}

\def \pasj {PASJ}

\def \mnras {MNRAS}

\title{Does the neutron star in Her~X-1 really show free precession?}

\ShortTitle{Her~X-1: free precession?}

\author{\speaker{R. Staubert}\thanks{presented by D. Klochkov on behalf of R. Staubert et al.}, D. Klochkov, D. Vasco\\
        Institut f\"ur Astronomie und Astrophysik, Universit\"at T\"ubingen, T\"ubingen, Germany\\
        E-mail: \email{staubert(klochkov,vasco)@astro.uni-tuebingen.de}}

\author{K. Postnov, N. Shakura\\
         Sternberg Astronomical Institute, Lomonossov University, Moscow, Russia\\
        E-mail: \email{kpostnov@gmail.com, nikolai.shakura@gmail.com}}

\author{R. Rothschild\\
         Center for Astrophysics and Space Sciences, Univ. of California San Diego, San Diego, USA\\
        E-mail: \email{rrothschild@ucsd.edu}}

\author{J. Wilms\\
         Dr. Remeis-Sternwarte Bamberg and Erlangen Center for Astroparticle Physics, Universit\"at Erlangen-N\"urnberg, Germany\\
        E-mail: \email{wilms@sternwarte.uni-erlangen.de}}

\abstract{The accreting X-ray pulsar Her X-1 shows two types of long-term variations, both with a period of $\sim$35\,d: 
1) A modulation of the flux with a ten day long \textsl{Main-On} and a 5\,d long \textsl{Short-On}, separated by two \textsl{Off}-states, 
and 2) A systematic variation of the shape of the 1.24\,s pulse profile. While there is general consensus that the flux 
modulation is due to variable shading of the X-ray emitting regions on the surface of the neutron star by the 
precessing accretion disk, the physical reason for the variation of the pulse profiles had remained controversial. 
Following the suggestion by Tr\"umper et al.~(1986)
that \textsl{free precession} of the neutron star may be responsible for the variation of the pulse profiles, we had developed 
physical models of strong feedback interaction between the neutron star and the accretion disk in order to 
explain the seemingly identical values for the periods of the two types of variations. In a deep analysis of 
pulse profiles observed by several different satellites 
over the last three decades we find now that the clock behind the pulse profile variations shows exactly the same
erratic behavior as the turn-on clock, even on short time scales ($\sim$100\,d), suggesting that there may in fact
be only one 35\,d clock in the system. If this is true, it presents a serious challenge for the idea of free precession 
of the neutron star and calls for alternative physical models for the variation in pulse shape.
}

\FullConference{8th INTEGRAL Workshop "The Restless Gamma-ray Universe" - Integral2010,\\
		September 27-30, 2010, Dublin, Ireland}

\begin{document}
\newpage

\begin{figure}
\begin{center}
\vspace{-3.0cm}
\includegraphics[bb=42 00 708 327,width=0.9\textwidth]{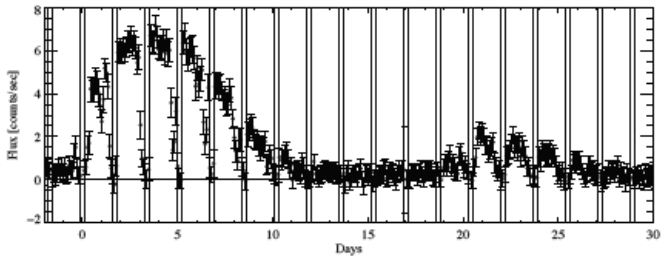} 
\vspace{-3mm}
\caption{Average 35\,d flux profile of Her X-1, generated by accumulating
light curves observed by \textsl{RXTE}/ASM for 35\,d turn-ons around binary
phase 0.2 (Fig.~3 of Klochkov et al. 2006). The vertical lines indicate the binary
eclipses.}
\end{center}
\end{figure}

\vspace{-2.5cm}

\section{35\,d flux modulation and turn-on history}
\vspace{-5.5mm}
Her~X-1 shows modulation of its X-ray flux on a $\sim 35$\,d period, with a "Main-On"
and a "Short-On", separated by two "Off" states. The sharp increase of the flux defines the
beginning of a cycle and is called "turn-on", it generally occurs either around binary phases 0.2 or 
0.7. Fig.~1 shows the mean flux profile for phase 0.2 turn-ons as observed by \textsl{RXTE}/ASM
(Klochkov et al., 2006).  The 35\,day modulation of the X-ray flux is generally explained 
by the precession of the accretion disk, that regularly blocks the view to the X-ray emitting 
regions near the magnetic poles of the neutron star.

The 35\,d turn-on clock is quite irregular, allowing the length of an individual cycle to be either
20.0$\times$P$_\mathrm{orb}$, 20.5$\times$P$_\mathrm{orb}$, or 21.0$\times$P$_\mathrm{orb}$ 
(Staubert et al., 2006) (with a small fraction of cases showing longer or shorter cycles). 
Adopting P$_{35}$ = 20.5$\times$P$_\mathrm{orb}$ (= 34.85\,d) as the mean ephemeris period, the 
turn-on history can be described by the so called ($O-C$)--diagram, which plots the difference 
between the observed turn-on and the calculated turn-on (using this ephemeris period) as
a function of cycle number (or time). Fig.~2 shows ($O-C$) since the discovery of Her~X-1 until 
today as a function of time in MJD. The two versions of the diagram (left and right) differ in the 
number of 35\,d cycles assumed to have occurred during the 602 day long Anomalous Low 
(AL3), which is centered at $\sim$\,MJD~51500. During this gap, the X-ray flux was very low 
because of the blocking by the low inclination accretion disk. For the interpretation of Fig.~2 we 
refer to Staubert et al. (2009), in the following called the "Two-clocks paper". In Fig.~2-left the 
gap corresponds to 18 cycles with a (short) mean duration of 19.7$\times$P$_\mathrm{orb}$, which
is probably the correct physical interpretation for a continued precession of the accretion
disk. In Fig.~2-right the gap corresponds to 17 cycles with a (long) mean duration of 
20.8$\times$P$_\mathrm{orb}$, which can be associated with a semi-regular clock with a mean
period of 20.8$\times$P$_\mathrm{orb}$ = 34.88\,d. In the "Two-clocks paper" (Staubert et al., 2009)
this period was associated with an underlying clock, viewed to be rather regular, namely
(free) precession of the neutron star, assumed to be responsible for the observed periodic 
variation of the shape of the pulse profiles. In this model, 
it was assumed, that the precession of the neutron star is the \textsl{master clock}
and that the precession of the accretion disk is locked to that of the neutron star by evidently
existing strong physical feed-back in the system, (nearly) synchronizing the periods of the 
"two clocks".

\begin{figure}
\begin{center}
\includegraphics[width=0.49\textwidth,angle=0]{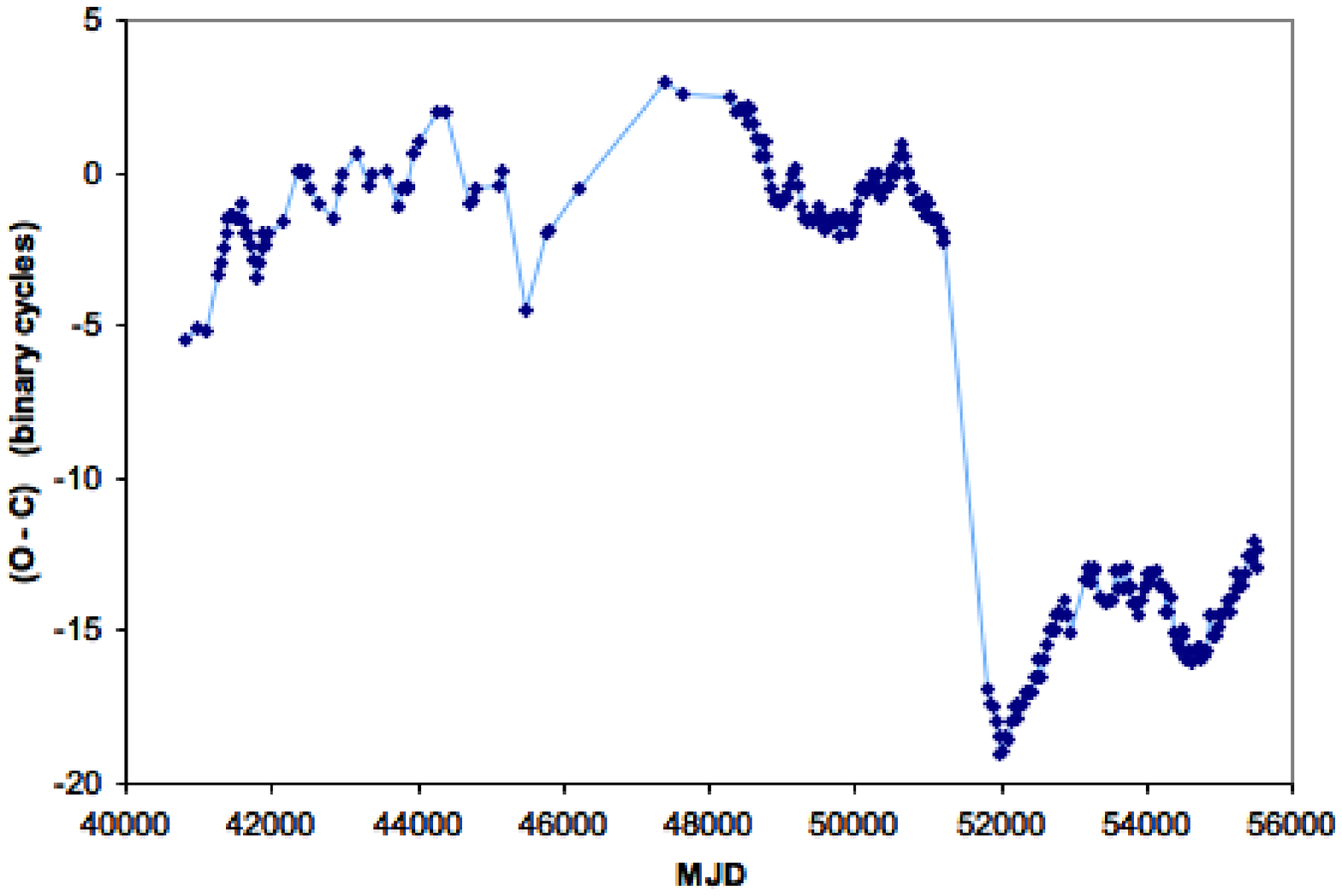} 
\hfill
\includegraphics[width=0.49\textwidth,angle=0]{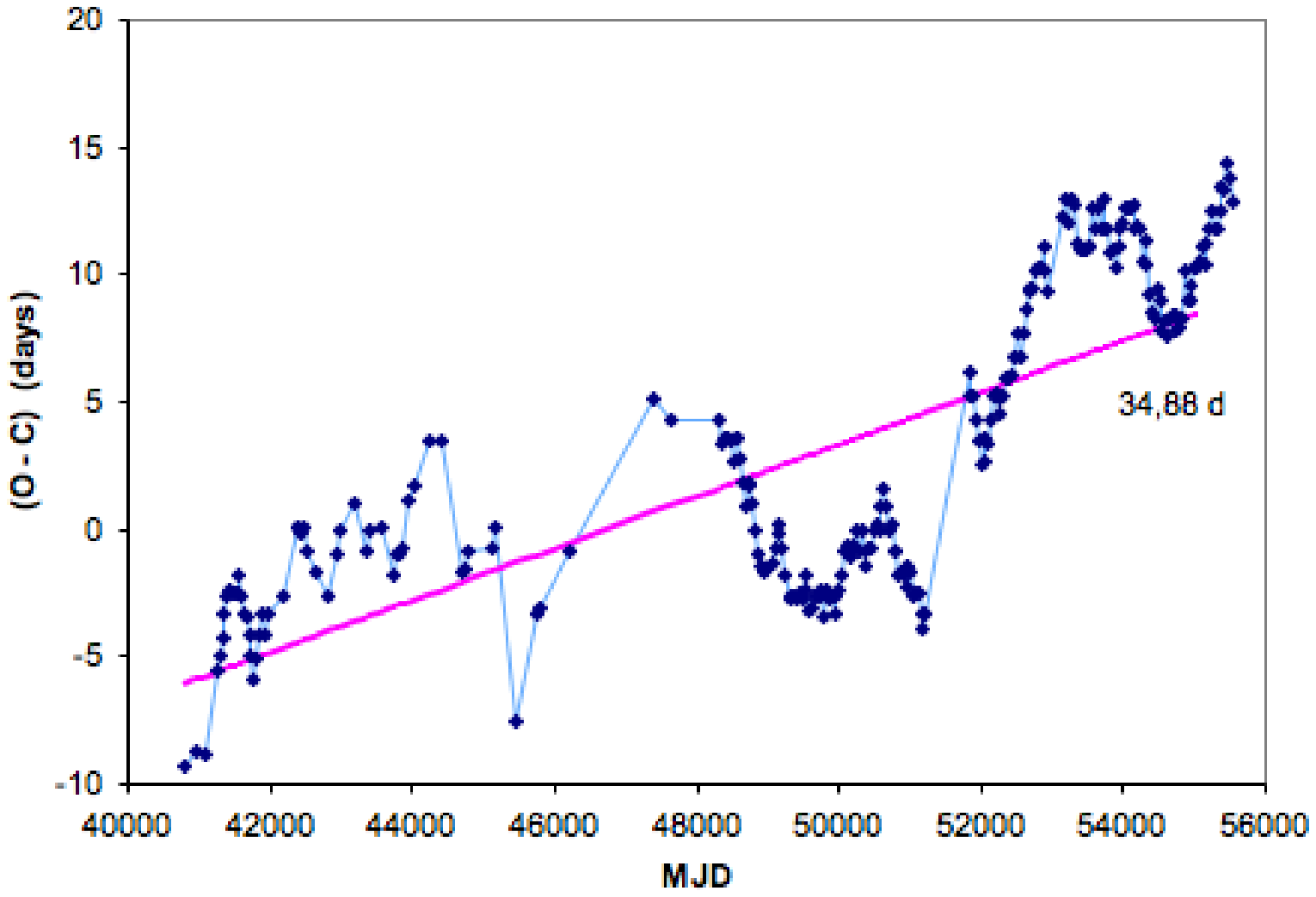} 
\caption{Turn-on history of Her~X-1: the ($O-C$)--diagram. \textsl{Left}: assuming that 
during the Anomalous Low around MJD 51500 (AL3) there were eighteen 35\,d
cycles. \textsl{Right}: the same assuming that there were only seventeen 35\,d cycles
(see Staubert et al., 2009 and text for further discussion).}
\end{center}
\end{figure}

\section{Pulse profile variations}

In observations of Her~X-1 by \textsl{EXOSAT} in 1984, Tr\"umper et al.~(1986)
discovered that the 1.24\,s X-ray pulse profiles vary in shape as a function of 
the phase of the 35\,d flux modulation. Observations by \textsl{Ginga} in 1989
and by \textsl{RXTE} starting in 1996 confirmed these findings and have added
a wealth of detailed information on the combined pulse shape and spectral
evolution of the pulsars beamed emission (e.g., Scott et al., 2000). Tr\"umper et al.~(1986) 
had suggested that the systematic variations in pulse shape is due to free
precession of the neutron star: the viewing angle towards the X-ray emitting polar cap 
region of the neutron star varies with the phase of the neutron star precession.
Shakura et al. (1998) applied a model of a precessing triaxial model to pulse
profiles of Her~X-1 observed by \textsl{HEAO-1}, Ketsaris et al. (2000) did so
for profiles observed by \textsl{RXTE}/PCA. 
Using all observations by \textsl{RXTE} from 1996 until 2005 we have verified
that the shape of the pulse profiles is reproduced every $\sim35$ days. 
A careful timing analysis was performed of all archived \textsl{RXTE} 
data on Her X-1 and pulse profiles were generated by folding with the measured 
pulse periods.  As an example, Fig.~3-left shows a set of pulse profiles (PCA, 9--13\,keV) 
for eight different 35\,d phases. The variation of the pulse shape is evident.
This systematic analysis has lead to the development of the "Two-Clocks-Model" 
(Staubert et al., 2009) and the successful modeling of the observed pulse profiles 
by a model of point- and ring-like emission from the polar caps of a neutron star 
with an offset-dipole field under the assumption of free precession (Postnov, 2004).

We have then started a model-independent investigation of the periodic pulse profile 
variations by constructing a template of those variations for the Her~X-1 "Main-On"
for photon energies 9--13\,keV. A set of observations were selected, providing a good 
coverage of the 35\,d phase range -0.05 to 0.15. This template contains flux normalized
pulse profiles for every 0.01 in phase. Any 9--13\,keV pulse profile observed during
a Main-On can then be compared to this template and the 35\,d phase can be determined 
(by $\chi^{2}_\mathrm{min}$ minimization). For the \textsl{RXTE} data we find that this is generally possible 
to an accuracy of  $\pm$0.02 in phase. First results of using this template with profiles
observed by \textsl{Ginga}, \textsl{RXTE} and \textsl{INTEGRAL} were presented
by Staubert et al. (2010), showing that the turn-on history and the history of variations
of the pulse profiles appeared to be strictly parallel, implying that the neutron star
precession (if it is indeed the reason behind the pulse profile variations) is synchronized
to the precession of the accretion disk. This raised the question, how the neutron star
would be able to significantly change its precessional period on rather short time
scales.

\begin{figure}
\begin{center}
\includegraphics[width=.48\textwidth]{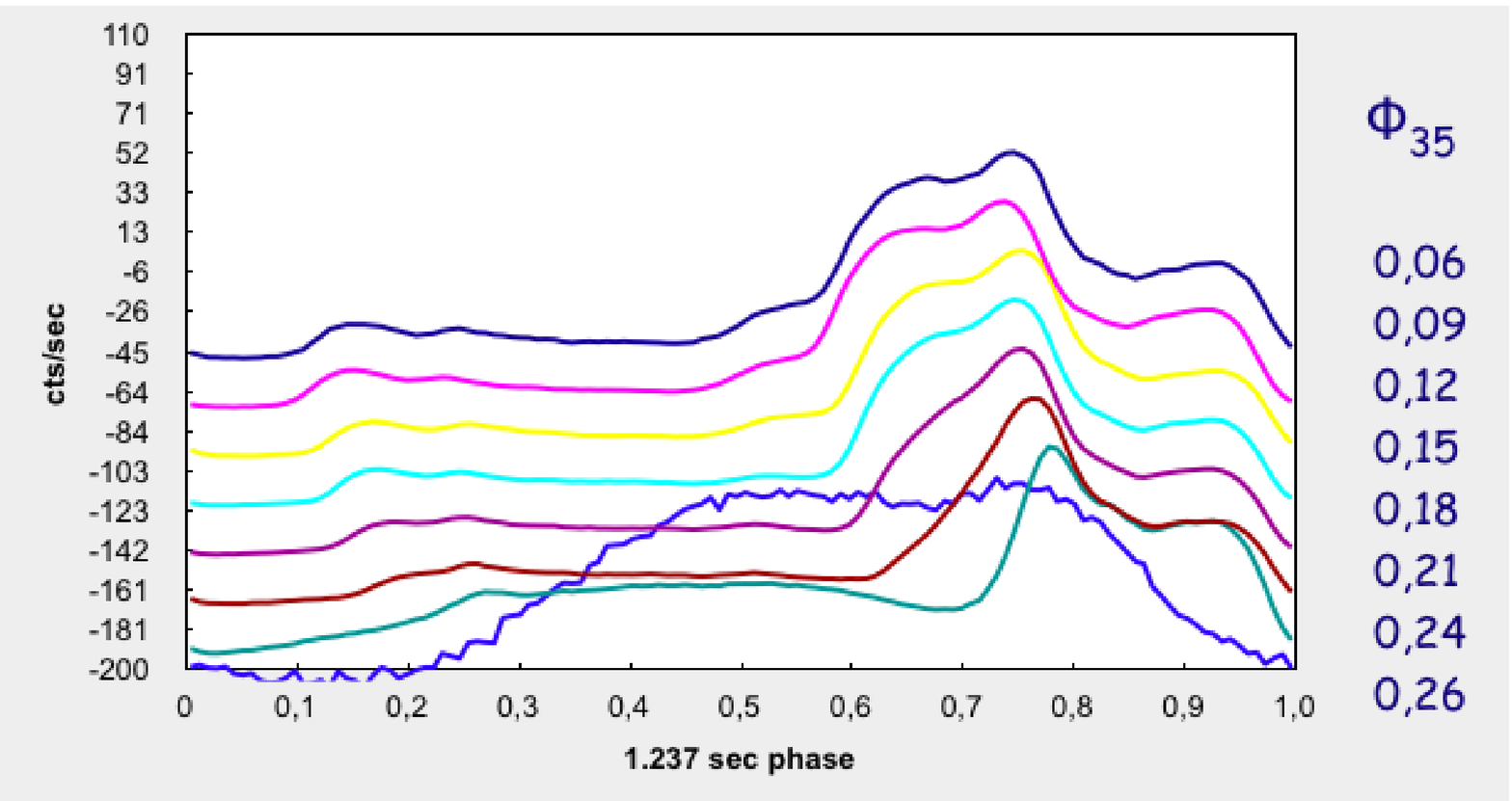} 
\hfill
\includegraphics[width=.49\textwidth]{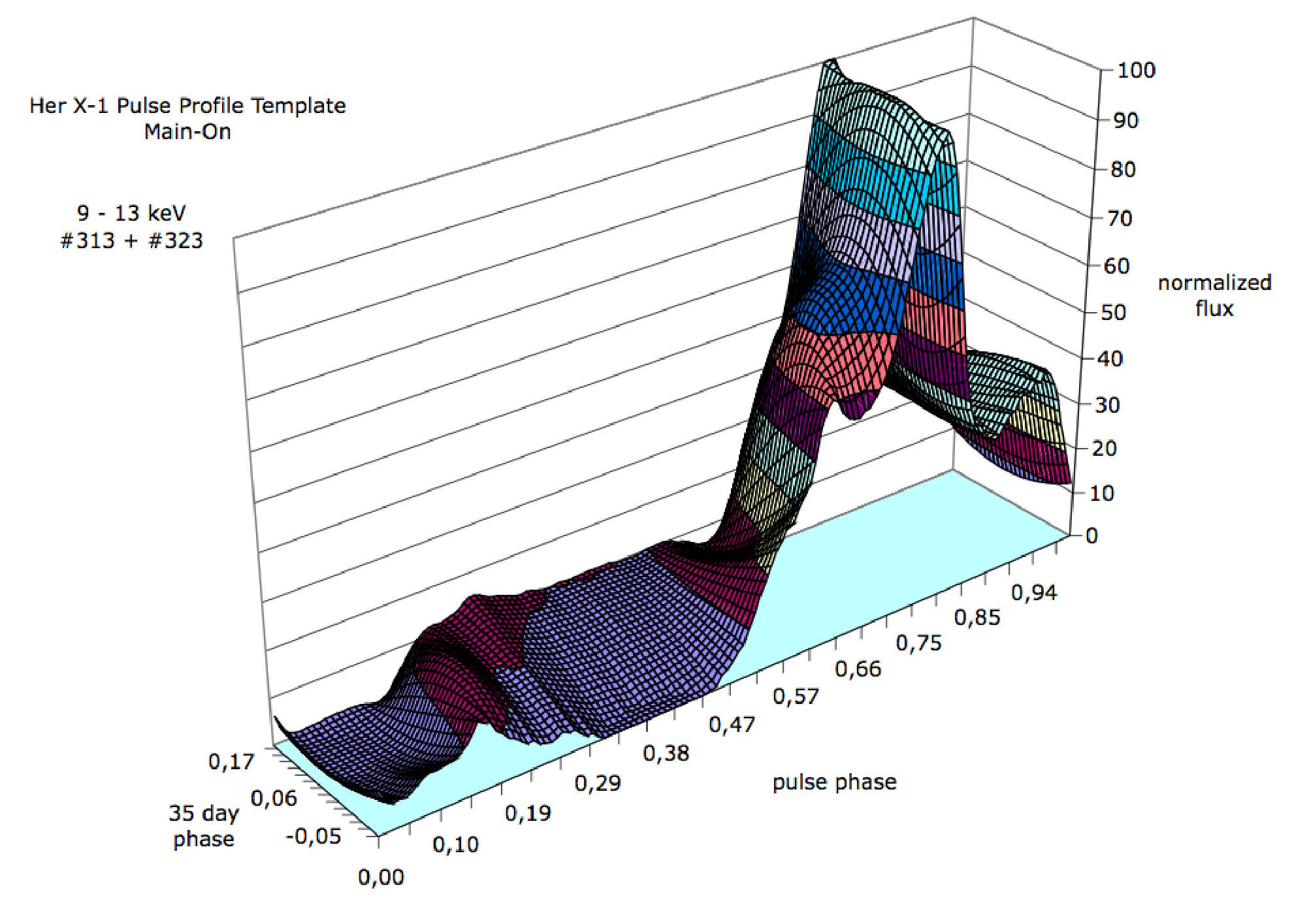} 
\caption{\textsl{Left}: Pulse profiles of Her~X-1 of 35\,day cycles of Dec 01 (cycle no. 313) and Nov 02 (cycle no. 323) 
as a function of 35\,day phase (RXTE, 9--13\,keV).
\textsl{Right}: 3D pulse profile template for the Her~X-1 Main-On (35\,d phase -0.05 to 0.22) of cycles
313 and 323 in the energy range 9--13\,keV. The resolution is 128 bins in pulse phase and 
28 bins in 35\,d phase.}
\end{center}
\end{figure}

We have since refined this method by constructing an extended template from \textsl{RXTE}/PCA
observations of two 35\,d cycles (nos. 313 and 323), which together provide a uniform coverage 
of a complete "Main-On" in the 35\,d phase range -0.05 to 0.22, again with a resolution of 0.01. 
The details of the construction of this template, its characteristics, and its usage are described 
by Staubert et al. (2011) (this Volume). A 3D-representation of the template (being a matrix of
128 bins in pulse phase by 28 bins in 35\,d phase) is given in Fig.~3-right.

\section{Results and discussion}

We compare the characteristics of the two 35\,d modulations (of the flux and of the variation in
pulse profiles) by plotting both into an ($O-C$)-diagram. For the flux, we use the turn-on, also
called the "accretion disk phase-zero" and plot the ($O-C$) as in Fig.~2-right. Correspondingly,
we determine a "pulse profile phase-zero" for all the Main-Ons for which there is at least one
observed pulse profile (for most Main-Ons there are several profiles). The "pulse profile 
phase-zero" is defined and calibrated using those cycles which were used to construct the template:
for cycles 313 and 323 both times of "phase-zero" are identical (by definition). For any other
cycle "pulse profile phase-zero" is found by determining the 35\,d phases of all available
profiles by comparison with the template and a subsequent linear extrapolation to phase zero 
using the fixed period of 34.85\,d (see Fig.~5-left of Staubert et al., 2011). 
Using pulse profiles of other Main-On cycles of Her~X-1 observed by \textsl{RXTE} 
over the last three decades plus one from \textsl{Ginga} and one from \textsl{INTEGRAL}, 
we find that these systematic variations are very stable and reproducible. 
The constructed template therefore allows to determine the 35\,d phase for any 
observed Main-On pulse profile in the 9--13\,keV range.

The observational result in the form of an ($O-C$)-diagram is summarized in Fig.~4. 
The green points (connected by the solid green line) represent the observed turn-on 
times (equal to "accretion disk phase-zero"). The magenta points represent the
times of "pulse profile phase-zero" as determined from the comparison of observed
pulse profiles with the pulse profile template and subsequent extrapolation to
phase zero. The observational evidence is quite clear: within statistical uncertainties,
the values for phase-zero as determined by the two different methods are identical,
and the "pulse profile clock" is just as irregular as the "turn-on clock".
Both clocks appear to be perfectly synchronized. 
We emphasize here, that the above result is completely model independent,
it is obtained using observational data only. However, the latest results from 
the continued effort in modeling the observed pulse profiles by a model 
assuming free precession of the neutron star (Postnov, 2004; a publication
is in preparation) leads to the same conclusion. This does suggest abandoning 
the concept of "two clocks" and to assume the existence of just \textsl{one 
underlying clock} which controls the variations of both zero-phases:
that of the turn-ons, and that of pulse profile phase zero.

What does the above result mean for the concept of (free) precession of the
neutron star in Her~X-1? We distinguish between two assumptions.

\begin{figure}
  \vspace{-15mm}
  \includegraphics[width=0.65\textwidth,angle=-90]{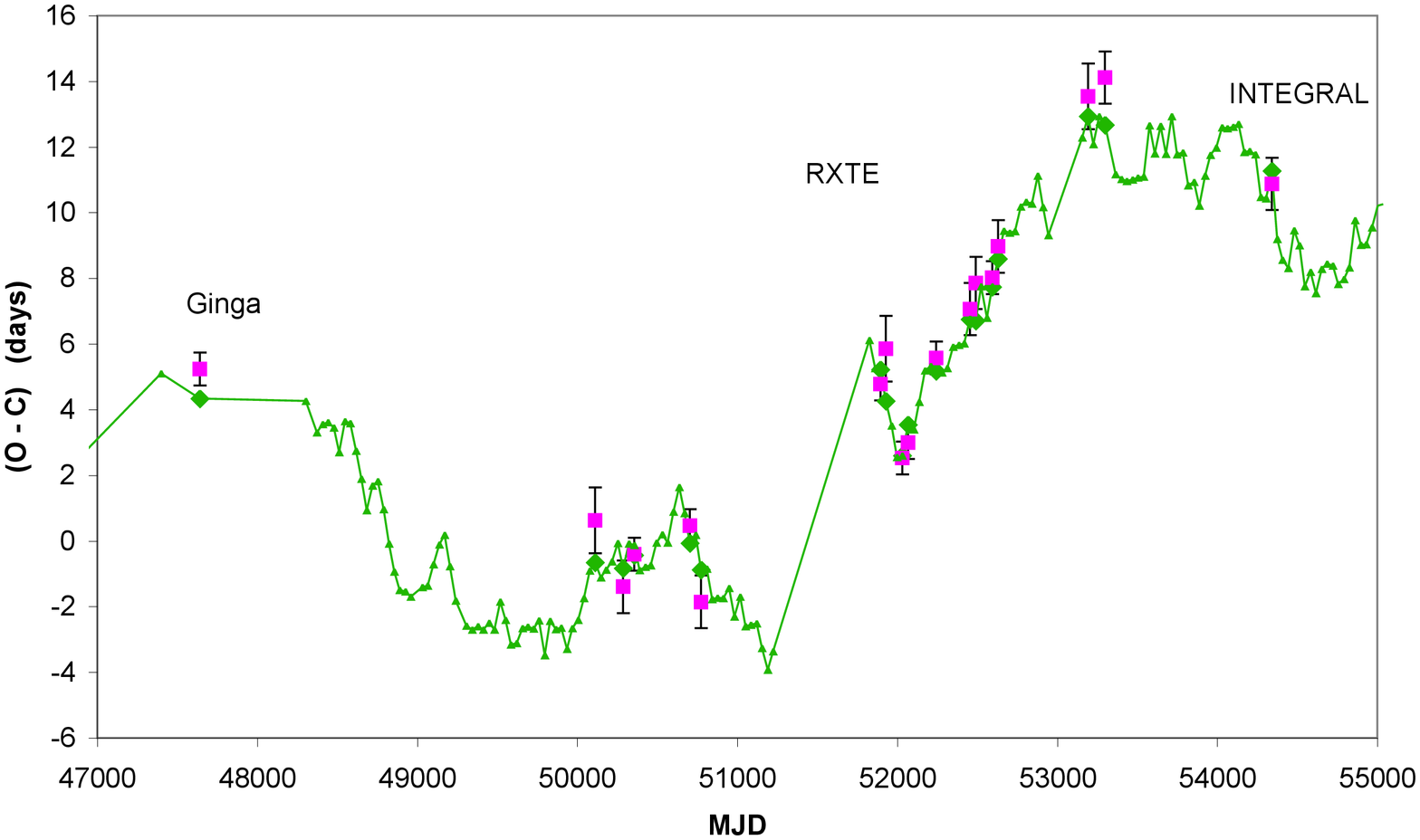}
  \vspace{-10mm}
  \caption{($O-C$) values in units of days for observed turn-ons (green) and for so far generated
                "pulse profile phase-zero" values from pulse profile fitting (mangenta),
                using profiles 9--13\,keV profiles for several Main-Ons observed by \textsl{RXTE}, plus 
                one from \textsl{Ginga} and one from \textsl{INTEGRAL}.}

   \label{Fig:fig4}
\end{figure}

First, assuming the precession of the neutron star does indeed exist and
is responsible for the variation of the pulse profiles and for those of the turn-on
times, we would then have to explain, how the neutron star could change its
precessional period by several percent on rather short time scales 
($\sim$100\,d). There seem to be no external forces that are strong enough to 
change the precessional period, e.g. by applying a torque to the principle
axis of inertia. The only possible origin could be inside the neutron star,
that is, if glitches occur or if the complex physics of the interior of a highly
magnetized neutron star with its crust and liquid core did allow for time
variable phenomena. For the relevance of free precession to our
understanding of matter at supra-nuclear densities, see e.g., Link (2007).
Equally difficult is the explanation of how the neutron star, if it is indeed the one 
master 35\,d clock in the system, would be able to transmit its precessional
motion to that one of the accretion disk (assuming that our long-term
concept is correct, that the turn-ons are due to the precession of the
accretion disk). We do indeed see strong feed-back in the binary system,
which has lead Staubert et al. (2009) to assume that the precession
of the neutron star may be the master clock in the system, under the
assumption, however, that the neutron star clock would be rather stable
and that the accretion disk would have "a life of its own" providing the
freedom to deviate from the strict clocking of the master - as observed
in the turn-on history. We now see, that the turn-ons and the pulse profiles
vary in strict synchronization, requiring extremely strong feed-back, the
existence of which appears questionable. Certainly, any neutron star
precession would be far from "free".

Second, assuming that there is no neutron star precession, we would
need a different explanation for the variation of the pulse profiles in
synchronization with the turn-ons. We would have to assume
that the precessional movement of the outer edge of the accretion
disk, thought to be responsible for the 35\,d flux modulation, is
mirrored at the inner edge of the accretion disk, and that its interaction 
with the magnetosphere of the neutron star and the way in which the accretion 
of matter proceeds along the field lines of the (probably) multi-pole magnetic 
field, is varying with precessional phase. Scott et al. (2000),
on the basis of \textsl{Ginga} and early \textsl{RXTE} observations,
proposed a model in which the changes in the shape and
spectral appearance of the pulse profiles are qualitatively explained by 
a combination of occultation of the X-ray emitting regions by a tilted and 
twisted precessing inner accretion disk and changes in the accretion 
geometry by the changing relative orientation between disk and neutron star 
magnetosphere. A detailed quantitative model would be needed to
verify this idea for the generation of the multi-component pulse profiles.

\end{document}